# Near-field perturbation of laser filament enabling simultaneous far-field THz diagnosis and broadband calculus processing

JIAYU ZHAO[1,*], YIFU TIAN[1], LINLIN YUAN[1], JIAJUN YANG[1], XIAOFENG LI[1], LI LAO[3], ALEXANDER SHKURINOV[4], YAN PENG[1,*], AND YIMING ZHU[1,2,*]

[1]*Terahertz Technology Innovation Research Institute, Terahertz Spectrum and Imaging Technology Cooperative Innovation Center, Shanghai Key Lab of Modern Optical System, University of Shanghai for Science and Technology, Shanghai 200093, China;*
[2]*Shanghai Institute of Intelligent Science and Technology, Tongji University, Shanghai 200092, China;*
[3]*Tera Aurora Electro-optics Technology Co., Ltd, Shanghai 200093, China.*
[4]*Faculty of Physics and International Laser Center, Lomonosov Moscow State University, Moscow 119991, Russia.*
*[zhaojiayu@usst.edu.cn](zhaojiayu@usst.edu.cn)
*[py@usst.edu.cn](py@usst.edu.cn)
*[ymzhu@usst.edu.cn](ymzhu@usst.edu.cn)

**Abstract:** Terahertz (THz) wave manipulation based on laser filaments—plasma channels formed by femtosecond laser-induced air ionization—has emerged as a promising platform for free-space THz applications. However, *in-situ* characterization of the spatially confined THz modes within filaments faces significant challenges due to the plasma's ultra-high intensity, which not only hinders direct near-field probing but also limits reliance on indirect far-field reconstruction. Here, we introduce a non-invasive near-field modulation scheme where a metal plate approaches the filament at submillimeter distances (comparable to THz wavelengths), perturbing the dielectric environment to convert the symmetric annular THz mode into an asymmetric state. This controlled transition enables far-field detection of broadband calculus behaviors (first- and second-order differentiation/integration) on time-domain THz waveforms and characteristic spectral transfer functions with $1/f$, $1/f^2$, $f$ or $f^2$ dependency (where $f$ is the THz frequency), thereby diagnosing the near-field THz mode confinement. Hence, the proposed approach synergizes near-field modulation efficiency with far-field detection robustness, advancing fundamental understanding of plasma-THz interactions and enabling novel all-optical signal processing for filament-based THz technologies.

## Introduction

In recent years, terahertz (THz) science has garnered widespread attention and undergone rapid development[1,2]. Particularly within the field of ultrafast optics, laser filaments—plasma channels generated via air ionization through femtosecond laser focusing—have become as a focal point in THz research. This interest stems not only from their function as broadband, high-intensity THz radiation sources[3,4], but also because laser filaments (and their arrays) inherently constitute a novel all-optical platform for free-space THz wave manipulation, offering broad application prospects and establishing them as an emerging research hotspot[5–12]. Recent studies have demonstrated diverse THz manipulation applications based on filament platforms, including single-filament-enabled THz enhancement, extinction, and filtering[5,6], THz imaging[13], and THz polarization control[14–16]; parallel dual-filament-driven broadband temporal-domain THz all-optical integration[7]; counter-propagating dual-filament-mediated far-field THz radiation envelope modulation[8,9]; orthogonal dual-filament-based THz super-resolution microscopic imaging[10,11]; and periodic multi-filament-array-guided free-space THz wave transmission[12]. Consequently, an emerging field dedicated to rapidly reconfigurable, air-medium-exclusive free-space THz manipulation is now under active development.

From these pioneering works, it becomes evident that the interaction between a single filament and THz waves constitutes the fundamental unit for constructing filament-THz application platforms (see also Supplementary Note 1). Existing theoretical studies reveal an annular THz electric field distribution at the cross-sectional periphery of laser filament, demonstrating the strong spatial confinement of THz waves[17]. This phenomenon represents a near-field effect occurring within the single-filament region at THz wavelength scales. Furthermore, the superposition of microscopic near-field THz modes between multiple filaments collectively governs macroscopic far-field THz radiation behaviors[7].

However, due to the ultra-high optical intensity within the filamentary region of air plasma and its inherently non-solid-state nature, conventional *in-situ* microscopic probing and direct experimental characterization (e.g., THz-SNOM[18]) of its near-field effects currently pose significant challenges for effective implementation. To address these issues, innovative quasi-near-field experimental approaches have been attempted. For instance, an ablation-resistant bakelite blade was employed to transversely intercept the filament-THz beam[17]. This method enabled detection of THz beam diameters smaller than their wavelengths within filament cross-sections, revealing a non-diffracting confined propagation state. However, the blade's insertion destabilized the filament structure, which degraded experimental signal-to-noise ratios (SNR). In reality, given that laser filamentation constitutes a highly complex and transient nonlinear optical process, direct physical contact between modulation devices and filaments should be avoided. Consequently, mature experimental methodologies for THz near-field characterization within filament regions remain underdeveloped.

As a compromise, researchers often resort to reconstructing near-field characteristics through far-field THz beam focusing and three-dimensional imaging near the focal zone[19-24]. While the observed low-diffraction transmission state[19,20] appears consistent with the spatial confinement effect of THz waves, the fundamental limitation of this far-field measurement approach arises because the original plasma filament medium no longer exists at the re-focused spot of the free-space THz beam, thereby inevitably struggling to accurately resolve near-field interactions between the filament and THz waves. For example in the literature[19-24], THz field distributions were typically reproduced at millimeter-size or larger spatial scales without near-field features.

In view of the above publications, an in-depth analysis of the near-field THz mode effect has not yet conducted, which limits the expansion of THz applications on laser filament platforms. The current key issue is how to operate as close to the filament as possible without being invasive and causing significant disturbances, and remain near-field operations rather than completely relying on the far-field THz beam reestablishment as in the literature. At the same time, to avoid the filament ablation effects of near-field probe, it is necessary to borrow the high SNR of mature far-field detection methods. A potentially feasible idea is to combine the near-field perturbation modulation and the far-field time/frequency-domain diagnosis, which is to precisely control the transition of the confined THz mode in the filament from one state to another in the near field, and then observe the characteristics of the changes between the two THz states in the far field, rather than just detecting a specific state itself.

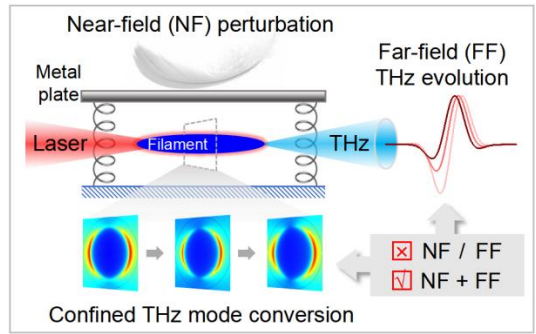

Fig. 1 | **The concept map of this work.** By applying near-field perturbations to the laser filament to alter the confined THz states within the filament region, one can ultimately detect characteristic changes in far-field THz signals.

In this work, as shown in Fig. 1, we propose a near-field perturbation of the laser filament combined with a far-field THz diagnostic scheme. Specifically, a metal plate is used to approach the filament parallelly to the order of THz wavelength, altering the dielectric constant distribution at the filament edge, thereby achieving the *in-situ* near-field modulation of the THz ring mode. Theoretically, such an operation will transform the originally symmetrically confined THz state into an asymmetric one within the filament. And this conversion process, in principle, leads to a broadband calculus evolution of the single-pulse THz time-domain waveform. Therefore, during this process, one only needs to monitor THz temporal signals in the far field and compare their dynamic changes with theoretical expectations. In this way, the near-field THz information can be reflected, thereby confirming the existence and conversion of confined THz modes within the filament region.

The advantages of this work can be preliminarily summarized as: (1) It combines the advantages of near-field and far-field methods, i.e., the former probes into the near-field region with high modulation efficiency, while the latter is technically mature with a high SNR, ultimately achieving the information of confined THz modes. And this scheme could also be extended to the detection of other physical quantities during the filamentation-based nonlinear process. (2) This work modulates the near-field THz state of the filament for the first time to the best of our knowledge, which is a new flexible control technology for plasma-based THz signal sources, capable of broadband first- and second-order calculus processing in the picosecond scale (please see Section 2 and 3). (3) Considering that the plasma filament is an electromagnetic (EM) waveguide for THz[12,17,25], microwaves[26-28], infrared[29], visible light[30-34], and even X-rays[35], our method of filament perturbation could serve as a prototype for extending all-optical calculus calculations throughout the EM spectrum. (4) In addition, micro-structures like periodic grooves can also be etched on the metal plate in Fig. 1, which could then perform additional operations such as THz filtering[36], thereby enriching the modulation means and applications of the filament-based THz platforms.

## Results

### Theory of THz mode conversion and calculus effect

Under the focusing of a lens, femtosecond laser pulses form a filamentary air-plasma channel near the focal point, which is referred to as a laser filament. And this filament can be simplified as a plasma column, whose radial plasma density exhibits a Gaussian distribution. The interaction between this plasma column and THz waves can be simulated by the COMSOL software[13]. Briefly, the simulation domain is set as a circle in diameter of 3 mm, which includes the cross-section of the plasma filament in the center with free electron density given by a two-dimensional Gaussian distribution function:

$$N_e = N_{e0}\exp\left(-\frac{x^2 + y^2}{b^2}\right) \quad (1)$$

where $N_{e0}$ represents the central free electron density set as $1.0\times10^{17}$ cm$^{-3}$, close to the value of optical filaments generated under typical laboratory laser power and focusing conditions. The parameter $b$ is set as 30 μm, representing the half-width at $1/e$ of the plasma filament column, corresponding to a full width at half maximum (FWHM) of about 50 μm. The dielectric constant of the plasma filament, denoted as $\varepsilon_r$, can be calculated by

$$\varepsilon_r = 1 - \frac{\omega_p^2}{\omega^2 - i\nu\omega} \quad (2)$$

where $\omega$ represents the THz angular frequency, and $\omega_p$ represents the plasma frequency given by

$$\omega_p = \sqrt{\frac{e^2}{m_e\varepsilon_0}N_e} \quad (3)$$

Here, $e$ represents the elementary charge of $1.60\times10^{-19}$ C, $m_e$ represents the effective mass of the electron of $9.11\times10^{-31}$ kg, and $\varepsilon_0$ represents the vacuum permittivity of $8.85\times10^{-12}$ F/m. $\nu \sim$ 1 THz represents the typical electron collision frequency within the filament.

The simulated THz mode distribution is shown in Fig. 2a at 0.3 THz as an example, inside which it can be observed that the THz electric field is concentrated at the radial edges of the filament, forming a spatially confined mode. This result shares similar properties with the analytical calculation given by the 1-D negative dielectric waveguide (1DND) model, as indicated in Supplementary Note 2. Due to the symmetry of this THz mode distribution, we use $E_S$ to characterize this state and also its maximum electric field strength, where $S$ stands for "symmetric".

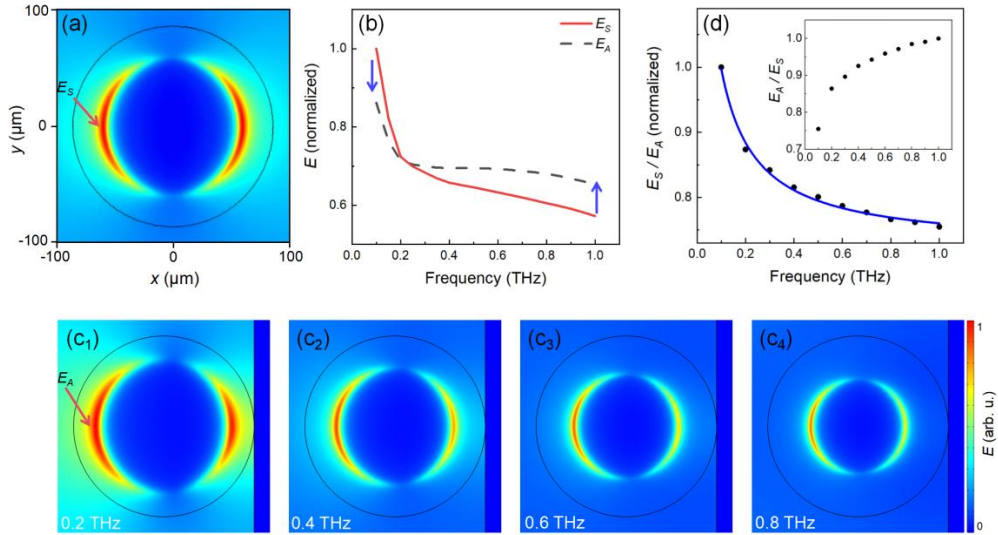

Fig. 2 | **Simulations of THz mode conversion during near-field perturbation on the filament. a** 0.3-THz electric field distribution in the symmetric state represented by $E_S$. **b** The variation of the maximum THz electric field as a function of THz frequency for both

symmetric and asymmetric states. **c₁-c₄** The THz electric field distributions in asymmetric states ($E_A$) at 0.2, 0.4, 0.6, and 0.8 THz, respectively. The metal plate is partly shown as the blue rectangle in the right side of **c₁-c₄**. **d** The ratio of the maximum THz electric field between symmetric and asymmetric states ($E_S/E_A$) as a function of THz frequency, with the blue fitting curve in a $1/f$ line type, where $f$ is the THz frequency. Inset: the result of $E_A/E_S$.

Next, simulations were conducted within the range of 0.1-1 THz, and variations of the peak THz electric field ($E_S$) is recorded as shown as the red line in Fig. 2b. One can see that as the THz frequency increases, $E_S$ generally exhibits a decreasing trend, which is consistent with Ref.[37], indicating that under spatial confinement, the amplitude of lower THz frequency components will be greater than those at higher frequencies. In this scenario, the question is how near-field perturbations could induce changes in this confined THz mode to achieve far-field detection of the state transitions? Considering the possible influence of polar materials on the THz electric field distribution, a metallic plate of aluminum with a width much larger than the filament diameter was introduced into the simulation domain. And this plate was made to approach the filament parallelly, until the final distance being on the order of the THz wavelength, in order to alter the dielectric constant distribution beside the filament and deform the initial THz mode.

Consequently, this arrangement indeed allows for the THz mode transitions from a symmetric distribution to an asymmetric one, with the electric field being stronger on the side away from the aluminum plate, as shown in Figs. 2c₁-c₄, corresponding to 0.2, 0.4, 0.6, and 0.8 THz, respectively. This asymmetrically confined THz state is denoted as $E_A$, where $A$ stands for "asymmetric." Similarly, we investigated the variation of $E_A$ from 0.1 to 1 THz, as shown by the black dashed line in Fig. 2b. It can be observed that although the variation trend of $E_A$ is similar to that of the symmetric state $E_S$, in detail $E_A$ is smaller than $E_S$ at lower frequencies and larger than $E_S$ at higher frequencies. This distribution characteristic can be understood as, due to the near-field proximity perturbation of the metallic plate, the original THz confinement by the filament is disrupted, making its feature of more low-frequency components and fewer high-frequency components in the $E(f)$ distribution not as pronounced as before (as pointed by arrows in Fig. 2b).

If we further quantify the ratio $E_S/E_A$, it is evident to decrease with the increasing THz frequency as shown as dots in Fig. 2d, which can be well fitted by a quasi-$1/f$ curve as the blue line with $f$ being the THz frequency. Conversely, the result of $E_A/E_S$ exhibits a increasing trend with $f$, as demonstrated in the inset of Fig. 2d. This $1/f$- or $f$-type ratio is, both mathematically and physically, a typical time-domain broadband integral-differential behavior as proved in Supplementary Note 3. Specifically, if we consider $E_S$ as the input signal, and the perturbed signal $E_A$ as the output one, then the conversion from $E_S$ to $E_A$ represents a differentiation of the single-pulse time-domain THz waveform; conversely, removing the perturbation of the metal plate represents a integration operation on the temporal THz signal (i.e., from $E_A$ to $E_S$).

Based on the theoretical analyses above, a correlation has been established between the near-field filament perturbation, the transformation of confined THz modes, and the changes in the far-field time- and frequency-domain THz radiation characteristics. Next, experiments were conducted to verify the integral-differential relationship of the THz temporal pulses, as well as the $1/f$ (or $f$) shaped envelope of the transfer function for the "input and output" THz spectra.

**The experimental interaction between the confined THz mode and the metal plate**

The experimental setup is schematically shown in Fig. 3 (see also Method and Supplementary

Note 4 for details). Briefly, a femtosecond laser was used with central wavelength of 800 nm, pulse width of 100 fs, repetition rate of 1 kHz, and single-pulse energy of 1.8 mJ. After focusing through a lens with focal length of 30 cm and a frequency-doubling crystal (BBO), the laser formed a two-color laser electric field, ionized the air around the focus and created a plasma filament, which subsequently radiated THz waves. Additionally, a Teflon plate was inserted after the filament to filter out the laser, and the transmitted THz pulses were finally detected by an electro-optic sampling (EOS) system.

In the experiment, we used an aluminum plate with a size matching the filament length (approximately 10 mm) but much larger than the filament diameter. The plate was placed parallel to the filament (below it) with an initial spacing of 10 mm, and then moved upward in steps of 0.2 mm. After each movement of the metal plate, we recorded the far-field time-domain THz waveform. Furthermore, when the plate was close to the filament (less than 5 mm), we reduced the movement step to 0.1 mm to more finely observe changes in the THz signals caused by the confined mode conversion. The minimum distance between the metal plate and the filament was approximately 0.2 mm, which not only ensured the interaction between the aluminum plate and the THz mode, but also avoided ablation of the plate by the filament.

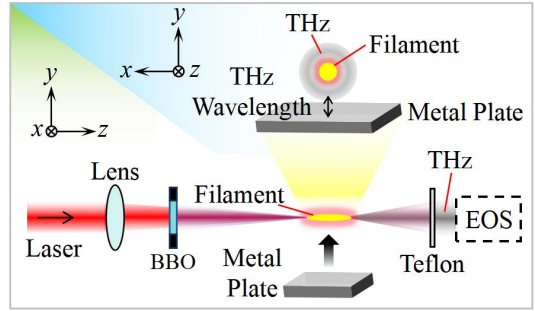

Fig. 3 | **Schematic diagram of the experimental setup for near-field filament perturbation and far-field THz detection.** The metal plate is parallelly approached to the filament until the distance between them is on the order of the THz wavelength. EOS is short for electro-optic sampling.

The recorded THz time-domain waveforms are shown in Fig. 4a. It can be observed that when the aluminum plate was relatively far from the filament, the THz signal did not exhibit significant changes. However, as the plate gradually approached the filament to a distance of $\Delta y$ = 1 mm or closer, the width of the THz temporal waveform became significantly narrower, and its peak experienced a noticeable advance in the time domain. This effect is more apparent in Fig. 4b, which shows the Fourier-transform (FT) frequency spectra corresponding to Fig. 4a, and a spectral peak shift towards higher frequencies occurs.

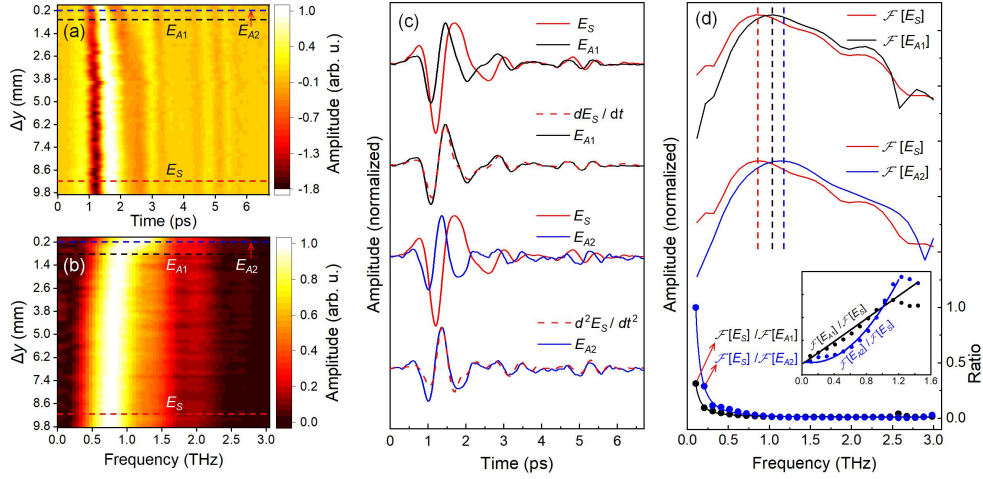

Fig. 4 | **Experimental far-field THz signals during near-field perturbation on the filament.**
**a** Time-domain THz signals as a function of the distance $\Delta y$ between the aluminum plate and the filament. The horizontal red, black, and blue dashed lines are located at $\Delta y = 9$ mm ($E_S$), 0.6 mm ($E_{A1}$), and 0.2 mm ($E_{A2}$), respectively. The three vertical dashed lines mark the fluctuations in the time domain caused by water vapor absorption of THz waves. **b** Fourier transform spectra of the THz time-domain signals in **a**, where the three horizontal dashed lines correspond to those in **a**. And the water vapor absorption of THz waves is omitted. **c** First and second derivatives of the $E_S$ signal from **a**, along with a comparison with the $E_{A1}$ and $E_{A2}$ signals. **d** Comparison of the extracted $E_S$, $E_{A1}$, and $E_{A2}$ spectra from **b**, along with the spectral ratios (black and blue dots), and the $1/f$, $1/f^2$, $f$ and $f^2$ shaped fitting lines (black and blue).

To verify the theoretically expected calculus relationship between these THz time-domain pulses, we extracted two signals as $E_S$ and $E_{A1}$ from Fig. 4a for comparison in the first row of Fig. 4c. The former is from the symmetric THz mode, which is located along the red dashed line in Fig. 4a at $\Delta y = 9$ mm. And the latter is from the asymmetric THz mode along the black dashed line at $\Delta y = 0.6$ mm. Then, $E_S$ was processed with differentiation and the result is shown in the second row of Fig. 4c together with $E_{A1}$. It can be seen that the differentiated $E_S$ (red dashed line) exhibits a high degree of similarity in temporal shape with $E_{A1}$ (black line). This result indicates that during the transition of the THz state from symmetric ($E_S$) to asymmetric ($E_{A1}$), the THz time-domain waveform undergoes a differentiation transformation; conversely, an integration effect would occur. These experimental results are consistent with our previous theoretical inference.

Additionally, we extracted a signal $E_{A2}$ along the blue dashed line at $\Delta y = 0.2$ mm from Fig. 4a and compared it with $E_S$ and its second-order differentiation, as shown in the third and fourth rows of Fig. 4c. Clearly, the two in the fourth row also show basic similarity in the time domain. This observation indicates that the near-field modulation depth of high- and low-frequency THz components increases when the metal plate is positioned closer to the filament, thus yielding higher (second) order differential/integral processing on the time-domain THz pulses. This further highlights the significant control potential of the proposed filament-perturbation method for manipulating THz signals.

Next, the THz spectra of $E_S$ and $E_{A1}$ from Fig. 4b were compared in Fig. 4d, where their ratio was also calculated. It can be seen that there is a typical distinction in their high- and low-frequency distributions: $E_S$ has stronger low-frequency components and a lower peak frequency (along the vertical red dashed line), while $E_{A1}$ has stronger high-frequency components and a higher peak frequency (along the vertical black dashed line). In this case, the ratio of the two spectra, i.e., $E_S/E_{A1}$ (or $E_{A1}/E_S$) in Fig. 4d, aligns well with the $1/f$ (or $f$) curve (black dots vs. black line). All above frequency-domain results are consistent with our theoretical expectations. Furthermore, similar results are also observed for $E_S$ (red line) and $E_{A2}$ (blue line) as shown in Fig. 4d. And their ratio $E_S/E_{A2}$ (or $E_{A2}/E_S$) shows a steeper decline (or growth) with the increasing THz frequency (blue dots vs. blue line), consistent with the trend of second-order calculus, i.e., $1/f^2$ (or $f^2$).

It is important to emphasize that, in Fig. 4a, there are three little peaks after the main THz pulse, which are time-domain fluctuations caused by the water vapor absorption of THz waves in air. Although their temporal positions show similar forward shifts with the main THz pulse as the distance $\Delta y$ decreases, their FT spectra (Supplementary Note 5) characterized by the water absorption lines (spectral dips) didn't change significantly during the near-field perturbation modulation on the filament. This is because the water vapor absorption effect always exists regardless of the interaction between the confined THz mode and the metal plate. It is also worth mentioning that in Figs. 4b and d, to avoid interference from the water absorption dips with the spectral structures, the aforementioned three little peaks in Fig. 4a were removed before performing the Fourier transform on the THz temporal signals. Therefore, no significant water absorption features can be observed in Figs. 4b and d. For this point, please see Supplementary Note 5 and 6.

**Substitution of the metal plate into a Teflon plate as the perturbator**

To further contrast the perturbation effect of polar (metallic) material on the filament and the transformation of the confined THz modes, we attempted to repeat the experiment by using a non-polar (Teflon) plate. Theoretically, due to the approximate transparency of non-polar materials to THz waves, they are not expected to significantly alter the THz dielectric constant distribution around the filament. Therefore, the THz state is predicted to remain unaffected, and the aforementioned differentiation/integration phenomena should not occur. This is actually what can be seen in Fig. 5, which has neither significant drift of the THz pulse in the time domain (Fig. 5a), nor obvious high-frequency enhancement trend in the spectra (Fig. 5b). All these observations agree with the theoretical expectations and further validate the applicability of the perturbation effect with metallic materials on the plasma filament and the transformation of the confined THz modes.

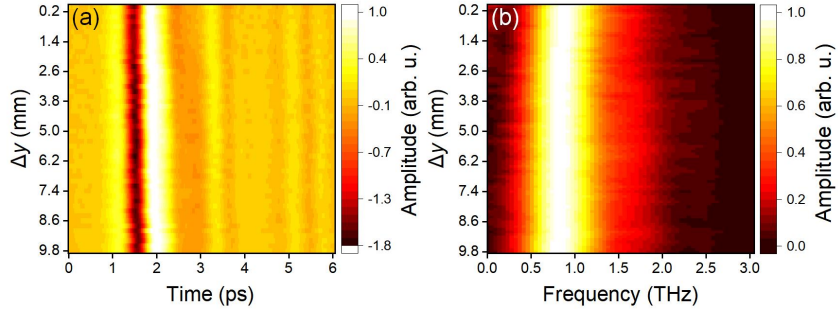

Fig. 5 | **Contrast experiment with a Teflon plate as the near-field perturbator on the filament. a** Time-domain THz signals as a function of the distance $\Delta y$ between the Teflon plate and the laser filament. **b** The corresponding frequency spectra for the signals shown in **a**, where the water vapor absorption of THz waves is omitted.

## Discussion

In summary, this work parallelly approaches a metal plate to the laser filament to a distance on the order of the THz wavelength, so as to interact with the confined THz mode at the periphery of the filament, thereby successfully changing the coupling state of the THz mode 0 from a symmetric one to an asymmetric one. At the same time, this near-field THz mode transition is manifested as the evolution of the first- and second-order calculus processing of the far-field time-domain THz waveforms, and the spectral amplitude ratio (i.e., the transfer function) satisfies a $1/f$, $1/f^2$, $f$ or $f^2$ line shape.

On one hand, the near-field THz perturbation within the filament region and the synergistic far-field THz time/frequency-domain radiation detection proposed in this work cleverly combine the advantages of near- and far-field methods: the former probes the near-field region with high modulation efficiency, while the latter is a mature technique with high SNR. The ultimately achieved transformation and diagnosis of the confined THz mode provide new insights into understanding and controlling the interaction between laser filaments and THz waves, paving the way for elucidating the related near-field mechanisms of plasma-THz coupling, guiding the precise manipulation of THz waves by novel laser filament platforms, and potentially extending to the detection of other physical quantities in nonlinear optical processes during laser filamentation.

On the other hand, the technique presented in this work represents a flexible method for manipulating THz signals based on the laser plasma stage. It enables all-optical computation of the first- and second-order temporal calculus of broadband THz pulses, where the laser plasma filament could serve dual roles as both the THz source and waveguide. And this calculus capability can be expanded to other EM bands beyond the THz range, such as microwaves[26-28], infrared[29], visible light[30-34], and even X-rays[35], since plasma filaments are capable of guiding these EM waves. Besides the displayed calculus processing, if periodic structures are further fabricated on the metal plate, functionalities such as single/multi-frequency THz filtering could be additionally realized[36], which easily enriches the modulation methods and application scenarios for THz wave manipulation based on the laser filament platforms.

## Method

### Experimental setup.

The THz time-domain spectroscopy (TDS) system used in this work consists of a femtosecond laser, from which the laser pulse is split into two paths. One was the pump beam and the other was the probe. The pump pulse is used to generate THz wave by means of laser plasma filament, which is created at the focus by focusing the pump beam through a converging lens and a BBO crystal. The exiting THz pulse from the filament was first collimated by an off-axis parabolic mirror, and then focused by another parabolic mirror onto the electro-optic sampling (EOS) setup. The probe beam was combined with THz pulse by a Pellicle beam splitter, performing THz measurement. The basic principle of THz-EOS is to slowly sample a fast THz transient with a delay-tunable probe pulse (through the delay line) by coherent detection in the time domain[38].

## Data availability

All data that support the findings of this study are present in the main text and the Supplementary Notes. All raw data generated during the current study are available from the corresponding authors upon request.

## Acknowledgements

The authors are grateful for the support from the National Natural Science Foundation of China (62588201); the National Key Research and Development Program (2023YFF0719200, 2023YFF0717900); the Shanghai Rising-Star Program (22QC1400300); and the 111 Project (D18014).

## Competing interests

The authors declare no competing interests.

## Additional information

**Supplementary information** The online version contains supplementary material available at xx